\documentstyle[prb,aps,epsf,floats]{revtex}
\begin{document}
\advance\textheight by 0.5in
\advance\topmargin by -0.25in
\draft

\twocolumn[\hsize\textwidth\columnwidth\hsize\csname@twocolumnfalse%
\endcsname

\title{Resonant Tunneling between quantum Hall states
at filling $\nu = 1$ and $\nu = 1/3$}
\author{C.L. Kane}
\address{Department of Physics, University of Pennsylvania\\
Philadelphia, Pennsylvania 19104 }

\date{\today}
\maketitle

\begin{abstract}
We study the problem of resonant tunneling between a Fermi liquid
and the edge of a $\nu=1/3$ fractional quantum Hall state.
In the limit of weak coupling, the system is adequately described
within the sequential tunneling approximation.  At low temperatures,
however, the system crosses over to a strong coupling phase,
which we analyze using renormalization group techniques.  We
find that at low temperatures a ``perfect" can be achieved by
tuning two parameters.  This resonance has a peak conductance 
$G = e^2/2h$ and is a realization of the weak backscattering limit
of the $g=1/2$ Luttinger liquid, as well as the non Fermi liquid
fixed point of the two channel Kondo problem.  We discuss
several regimes which may be experimentally accessible as
well as the implications for resonances recently observed in
cleaved edge overgrowth structures.

\end{abstract}
]

\section{Introduction}

The rich structure of the fractional quantum Hall effect 
offers a controlled laboratory for studying the physics
of strong correlation.  
As first argued by Wen\cite{wen}, the low energy excitations
at the edge of a quantum Hall state form a novel quantum
fluid known as a chiral Luttinger liquid.  
This theory has led to a number of predictions regarding
the effects of tunneling and resonant tunneling
between
quantum Hall states\cite{wen,kf1,moon}.  Experiments by Milliken Umbach
and Webb\cite{webb} probed tunneling and resonant tunneling at a 
quantum point contact. They observed a striking difference
in behavior at filling factors $\nu = 1$ and $\nu = 1/3$,
which was consistent with these predictions.
More recent experiments on a cleaved edge overgrowth
structure by Chang, Pfeiffer and West\cite{chang1} observed
the predicted power law scaling of the tunneling current
with voltage and temperature
at $\nu = 1/3$ over several decades of current.
While questions remain about other filling factors\cite{grayson}, 
the agreement between experiment and theory has been quite encouraging
for the Laughlin states at $\nu = 1$ and $1/3$.

In this paper we analyze the problem of resonant tunneling
between two different quantum Hall states, $\nu = 1$ and 1/3
- or equivalently between a Fermi liquid and $\nu = 1/3$.
Our motivation is twofold.  First, in their recent cleaved
edge overgrowth experiments, Grayson et al. have observed
interesting resonances in tunneling between a metal and 
$\nu = 1/3$\cite{grayson2}.  Second, this problem is interesting from a
theoretical point of view because, as we explain below, 
a ``perfect" resonance in this problem constitutes a realization
of the perfectly transmitting fixed point of a $g=1/2$ Luttinger
liquid, which is equivalent to the non Fermi liquid fixed point
of the two channel Kondo problem.

Our analysis is closely related to that of resonant
tunneling in a $\nu = 1/3$ quantum point contact, which
is closely related to the problem of tunneling through a barrier
in a $g=1/3$ Luttinger liquid\cite{kf2}.  
The limit of strong pinch off corresponds to a barrier with very
small transmission, whereas the limit weak pinch off corresponds
to a small barrier which backscatters only weakly.  As
argued in Ref. \onlinecite{moon} a perfect resonance is controlled
by the perfectly transmitting fixed point.
Such a resonance
has a peak conductance of $e^2/3h$ and a universal lineshape
which is determined by the crossover between the perfectly
transmitting and perfectly reflecting fixed points.  

The problem of tunneling between $\nu = 1$ and $\nu = 1/3$ can be
mapped onto that of tunneling through a barrier
in a $g = 1/2$ Luttinger liquid.  
This has led to the interesting suggestion that
the interface between $\nu = 1$ and $\nu = 1/3$ may 
allow for the realization of the perfectly transmitting
limit of the $g=1/2$ Luttinger liquid\cite{chamon1,chklovskii}. 
Such a state would have interesting properties.  Since
the ``perfect" two terminal conductance $e^2/2h$ is larger than
the conductance of the quantum Hall state, $e^2/3h$
it would act 
as a DC step-up transformer\cite{chklovskii}.  In addition,
the weak backscattering in this state would occur via quasiparticles 
with charge $e/2$, which could in principle be detected in a shot noise 
experiment\cite{sandler}.

However, realizing this state presents a subtle problem.
Unlike the $\nu = 1/3$ point contact, where one can 
physically separate the counter
moving edge states, there is no direct way of ``engineering"
the $g=1/2$ perfectly transmitting fixed point.
In principle it could be achieved using the
adiabatic contact, suggested by Chklovskii and Halperin
\cite{chklovskii},
in which the quantum Hall fluids
are connected smoothly by a quantum wire, in which the
Luttinger parameter $g$ can vary continuously.  
Recently, Sandler et al.\cite{sandler} have argued that 
the perfectly transmitting fixed point may be
approached by scaling to sufficiently {\it high} energies.  
However, since there are many {\it irrelevant} operators at
this fixed point, it is unlikely that the system would flow near 
the fixed point without tuning several parameters.

In this paper, we show that the perfectly transmitting limit can
be reached by resonantly tunneling through an impurity state.
A ``perfect" resonance can be reached at zero temperature
by tuning two parameters.  Such a resonance has a peak
two terminal conductance of $e^2/2h$ and has a universal
temperature dependent
lineshape which has been computed in the context of a 
$g=1/2$ luttinger liquid\cite{kf2}.  The remainder of the paper
is organized as follows.  After describing the model, we
discuss the resonances in the weak tunneling limit, when the
tunneling is sequential.  Then using a renormalization group
analysis we study the crossover to the perfect resonance,
Finally we discuss
the experimental implications of our results and the connection
with the recent observations by Grayson et al.\cite{grayson2}

\section{The Luttinger Liquid Model}

We begin with the chiral Luttinger liquid model describing
the coupling between the $\nu = 1/3$ and $\nu = 1$
edge states and the impurity state.  
The Hamiltonian may be written, 
$ H = H_0^{1} + H_0^{3} + H_{\rm imp} + H_T^{1} + H_T^{3}$.
The edge states are described by
\begin{equation}
H_0^m = \int dx u_m (\partial_x \phi_m)^2
\end{equation}
where the fields $\phi_m$ satisfy the Kac Moody algebra
\begin{equation}
[\phi_m(x'),\partial_x\phi_m(x)] = (2\pi i/m) \delta(x-x'),
\end{equation}
and $u_m$ determines the velocity of propagation.
We consider a single impurity state with energy $\varepsilon_0$,
\begin{equation}
H_{\rm imp} = \varepsilon_0 d^\dagger d,
\end{equation}
where $d^\dagger$ creates an electron in that state.
The tunneling of electrons between the impurity state and
the edges is described by
\begin{equation}
H_T^m = t_m \tau_c^{-1} d^\dagger e^{i m \phi_m}  + {\rm h.c.}
\end{equation}
where $t_1$  and $t_3$ are dimensionless tunneling amplitudes and
$\tau_c$ is a short time cutoff.

Away from a resonance, when $\varepsilon_0 \ne 0$, electrons
may tunnel virtually through the impurity.  This leads to a 
nonresonant conductance which vanishes as $T^2$ due to the power 
law tunneling density of states of the $\nu = 1/3$ edge.  However 
when $\varepsilon_0$ is tuned through zero, the energy denominator
for the virtual state becomes small, and there can be resonant
transmission through the impurity state.  
We will analyze the resonances first in the perturbative 
sequential tunneling limit, and then perform a renormalization
group analysis, which can describe the perfect resonance.

\section{Sequential Tunneling Limit}

In the limit, $t_m \ll 1$, tunneling is so infrequent that
successive hops onto and off of the impurity will be
uncorrelated.  This ``sequential tunneling" regime
has been studied in detail in the
context of Luttinger liquid theory by Chamon and Wen\cite{chamonwen} and by
Furusaki and Nagaosa\cite{furusaki}.  The result of this analysis is that
the tunneling conductance is given by
\begin{equation}
G = {\pi\over 2}{e^2\over h}
{\Gamma_3(\varepsilon_0,T) \Gamma_1(\varepsilon_0,T) 
\over {\Gamma_3(\varepsilon_0,T)+\Gamma_1(\varepsilon_0,T)}}
{1\over T} {\rm sech}^2 {\varepsilon_0\over {2T}}.
\end{equation}
$\Gamma_m(\varepsilon,T)$
is the inverse lifetime for tunneling into the $\nu = 1/m$ lead from the impurity,
which reflects the power law dependence of the tunneling density of states,
\begin{equation}
\rho(\varepsilon) \propto \varepsilon^{2\Delta^0_m -1}.
\end{equation}
Here $\Delta_m^0$ is the bare scaling dimension of the electron
tunneling operator, given by
\begin{equation}
\begin{array}{rcl}
\Delta_1^0 &=& 1/2 \\
\Delta_3^0 &=& 3/2.
\end{array}\end{equation}
In the following section we will see that $\Delta_m$ can be renormalized.
We therefore compute the lifetimes for general $\Delta_m$,
\begin{equation}
\Gamma_m(\varepsilon,T) = 2\pi t_m^2 \tau_c^{-1}(2\pi T \tau_c)^{2\Delta_m -1}
F_m({\varepsilon\over{2\pi T}}),
\end{equation}
with
\begin{equation}
F_m(x) = 2 {\rm cosh} \pi x 
{|\Gamma(\Delta_m + ix)|^2 \over {\Gamma(2\Delta_m)}}.
\end{equation}

In the perturbative regime, in which $\Delta_m$ are given by (7), this
formula reduces to
\begin{equation}
\begin{array}{rcl}
\Gamma_1(\varepsilon,T) &=& c_1 \tau_c^{-1} t_1^2 \\
\Gamma_3(\varepsilon,T) &=& c_3 \tau_c t_3^2 \left(
\varepsilon^2 + \pi ^2 T^2 
\right)
\end{array}
\end{equation}
where $c_1$ and $c_3$ are dimensionless constants of order unity.  
It is useful to distinguish the following limiting cases:

(1) When $\Gamma_1 \ll \Gamma_3$, the conductance is limited
by the tunneling into the $\nu = 1$ edge.
In this regime, 
\begin{equation}
G = {\pi\over 2} {e^2\over h}   {\Gamma_1 \over T} {\rm sech}^2{\varepsilon_0\over {2T}},
\end{equation}
and the resonances are Fermi liquid like, with a peak conductance
scaling as $1/T$ and a lineshape given by the derivative of the
Fermi function.

(2) When $\Gamma_3 \ll \Gamma_1$ the conductance is limited by
tunneling into the $\nu = 1/3$ edge.  In this case,
\begin{equation}
G \propto {\varepsilon_0^2 + \pi^2 T^2 \over T} 
{\rm sech}^2{\varepsilon_0\over{2T}}.
\end{equation}
Thus, the peak conductance decreases as $T$ as the temperature is lowered
in contrast to the $T^2$ behavior off resonance.
The lineshape is slightly modified from the derivative of the Fermi
function.  

\section{Renormalization Group Analysis}

It is important to emphasize that the sequential tunneling approximation
is valid only when tunneling is uncorrelated.  This will be true
provided $\Gamma_1, \Gamma_3  \ll T$.  To go beyond this
approximation it is useful to develop a perturbative
renormalization group analysis. 
We will now focus on the resonance peaks, when
$\varepsilon_0=0$.  To leading
order, the scaling equations for $t_1$ and $t_3$ are
given by 
\begin{equation}
{d t_m \over {d\ell}} = ( 1 - \Delta_m) t_m.
\end{equation}

To lowest order in $t_m$ $\Delta_m$ is simply given by (7).
Thus, to leading order, $t_3$ is irrelevant, while $t_1$ is
relevant.  The scaling behavior of $\Gamma_m$ in
(8) follows directly,
\begin{equation}
\Gamma_m \sim (t_m^R)^2 T \propto T^{2\Delta_m-1}.
\end{equation} 
where $t_m^R$ is the renormalized value of $t_m$ at an energy
scale of order $T$.  The perturbative analysis is then
valid provided $t_m^R \ll 1$, or equivalently $\Gamma_m \ll T$.

For finite $t_m$, $\Delta_m$ is renormalized.
This effect has been worked out in detail by the author in
Appendix B of Ref. \onlinecite{kf2} in the context of resonant
tunneling in a Luttinger liquid.  Here we generalize those results
to the problem of tunneling between $\nu = 1$ and $\nu = 1/3$.
The resulting flow equations for $\Delta_m$ are
\begin{equation}
\nonumber
{d\Delta_1\over{d\ell}} = - 4 \Delta_1 t_1^2 - 2(\Delta_1+
\Delta_3 -2) t_3^2\\
\end{equation}
\begin{equation}
{d\Delta_3\over{d\ell}} = - 4 \Delta_3 t_3^2 - 2(\Delta_1+
\Delta_3 -2) t_1^2
\end{equation}
It is instructive to analyze the renormalization group flows
in limiting cases.  

(1) For $t_3 = 0$, the resonant
state is coupled only to the $\nu = 1$ edge.   In this case,
$t_1$ is relevant and the system flows to a strongly coupled
phase with $t_1 \rightarrow \infty$, $\Delta_1 = 0$ and $\Delta_3 = 2$.
Since $\Delta_3 = 2$, $t_3$ is an irrelevant perturbation in
this phase.  The tunneling conductance can then be found
perturbatively to vary as
$G \propto t_3^2 T^2$.
This is the same as non resonant tunneling
between $\nu = 1$ and $\nu = 1/3$.
In this phase the impurity has effectively merged with the $\nu = 1$
quantum Hall fluid.  

\begin{figure}
   \epsfxsize=3in
   \centerline{\epsffile{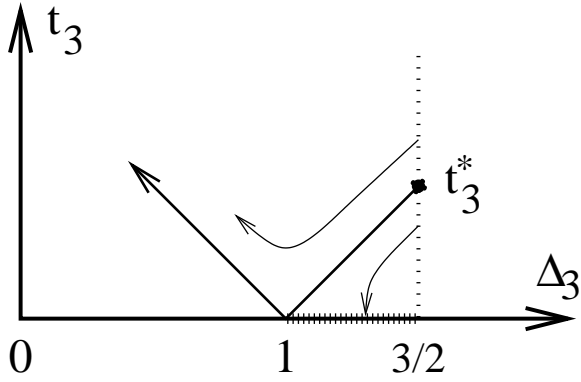}}
\caption{Renormalization group flow diagram as a function
of $t_3$ and $\Delta_3$ for $t_1=0$.  For $t_3<t_3^*$
$t_3$ flows to zero, and the impurity state decouples from
the $\nu = 1/3$ state.  For $t_3>t_3^*$ the impurity
state merges with the $\nu = 1/3$ state.}
\end{figure}

(2) For $t_1 = 0$, the resonant state is coupled only to the
$\nu = 1/3$ edge.  To leading order $t_3$ is irrelevant.  However,
$\Delta_3$ is renormalized downward from its initial value of
$3/2$.  Thus, as indicated in Fig. 1,
when $t_3$ reaches a critical value $t_3^* \approx 0.22$ the system flows
to a Kosterlitz-Thouless like 
fixed point at $t_3=0$, $\Delta_3 = 1$. 
For $t_3 > t_3^*$ the system flows to a strongly coupled phase
with $t_3 \rightarrow \infty$, $\Delta_3 = 0$ and $\Delta_1 = 2$.
Here, the impurity has merged with the $\nu=1/3$ quantum Hall
fluid.  Treating $t_1$ perturbatively again leads to 
a tunneling conductance varying as $T^2$.
On the other hand, for $t_3<t_3^*$, $t_3$ flows to zero with
a modified exponent: $1 < \Delta_3 < 3/2$.  In this
case $\Gamma_3(T) \propto T^{2\Delta_3 -2}$.
Since $t_1$ remains relevant, however, this state
is  unstable for finite $t_1$ and at low temperatures
the impurity merges with the $\nu = 1$ quantum Hall fluid.

When both $t_1$ and $t_3$ are finite it is clear that there
are two phases: the impurity merges with either the
$\nu = 1$ or the $\nu = 1/3$ quantum Hall fluids.  For fixed
$t_1$ as $t_3$ is increased the system undergoes a transition
between these two phases, as indicated in the phase diagram
Fig. 2.  Integration of the flow equations
shows that precisely at this transition, 
$t_1$ and $t_3$ are equal and grow to infinity, with 
$\Delta_1 = \Delta_3 = 1/2$.  We will analyze this case in the
following section.

\begin{figure}
   \epsfxsize=3in
   \centerline{\epsffile{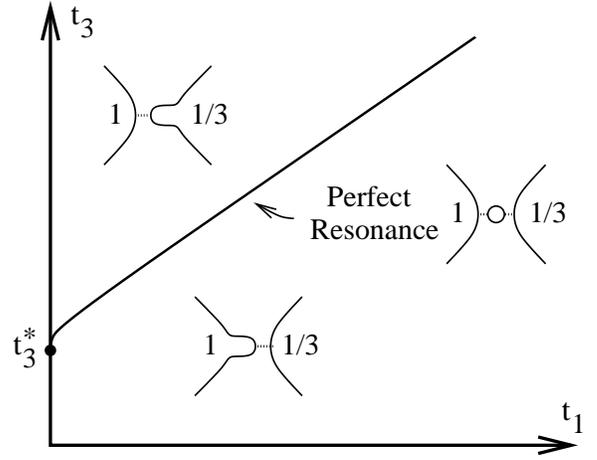}}
\caption{Zero temperature phase diagram for an impurity resonantly coupled to 
$\nu=1$ and $\nu=1/3$ states as a function of the
couplings $t_1$ and $t_3$.  For large $t_1$ the impurity
merges with the $\nu = 1$ state, while for large $t_3$
the impurity merges with the $\nu = 1/3$ state.
At the transition between the two phases is the perfect
resonance, which is described by the perfectly
transmitting fixed point of a Luttinger liquid at $g=1/2$.}
\end{figure}

\section{The Perfect Resonance}

The fixed point describing
the phase transition is not accessible within this perturbative analysis.  
However, when $\Delta_1 = \Delta_3$ and $t_1 = t_3$, this problem
is equivalent to resonant tunneling between two $g=1/2$ Luttinger
liquids in the limit in which the two tunneling barriers are symmetric.
As argued in Ref. \onlinecite{kf2}, this is precisely the condition
for a ``perfect resonance", where the system flows to the perfectly
transmitting fixed point.  

Equivalently, this model is can be mapped to the anisotropic two channel
Kondo problem\cite{matveev,yi}.  
In this mapping, the occupation of the impurity state corresponds
to the state of the Kondo spin, and the two leads correspond to
the two channels.  The tunneling matrix elements $t_m$ 
play the role of the transverse couplings $J_\perp^i$ for the
two channels, $i$.  The scaling dimensions $\Delta_m$ are related 
to $J_z^i$.   When $\Delta_1 = \Delta_3 = 1/2$, the Kondo
model is in the Toulouse limit. 

Under this mapping, the two phases in which the impurity state merges
with either $\nu = 1$ or $\nu = 1/3$  can easily be identified.  
When there is channel anisotropy in the two channel Kondo problem
the Kondo spin tends to form a singlet with one or the other
channels.  As the couplings are varied, there is a transition in
which the Kondo spin switches its allegiance.  Precisely at this
transition, which occurs when the channels are symmetric, the 
system flows to the non Fermi liquid fixed point in which the
spin is shared between the two channels.

Having established that the resonance fixed point is equivalent
to those of both the perfectly transmitting $g=1/2$ Luttinger liquid and
the two channel Kondo problem, we may now describe its properties.
Precisely on resonance, the conductance is $G = e^2/2h$.
Slightly away from the resonance, the lineshape
is controlled by the single relevant operator at the resonant
fixed point.  This operator describes equivalently $2k_F$ backscattering in 
the Luttinger liquid model, or channel anisotropy in the Kondo model.
This leads to a resonance with a temperature dependent
width proportional to $T^{1/2}$ and a universal lineshape
\cite{kf2}.
\begin{equation}
G(T,\delta) = \tilde G(\delta/T^{1/2})
\end{equation}
where $\delta$ is the tuning parameter for the resonance and
\begin{equation}
\tilde G(X) = {e^2\over {2h}} \int_{-\infty}^\infty dy
{e^y\over{(e^y+1)^2}}{y^2\over{y^2+X^4}}.
\end{equation}

\section{Experimental Implications}

The resonances depend critically on temperature as well
as the coupling between the impurity state and the
two leads.  Here we outline the various possible regimes,
depending on these parameters.  It is most useful to 
characterize the coupling to the leads by the 
inverse lifetimes $\Gamma_1$ and $\Gamma_3$, though one must
keep in mind that they can depend on temperature.  
In the limit of weak coupling, $\Gamma_1$ is temperature
independent, while $\Gamma_3 \propto T^2$.  For stronger coupling
the temperature dependence is altered.

If $T \gg \Gamma_1,\Gamma_3$,
then tunneling is sequential, and the resonances should be accurately
described by equation (5).   There are three distinct cases
depending on $\Gamma_1$ and $\Gamma_3$:

(1) {\it Fermi Liquid dominated sequential tunneling}  
If $\Gamma_1 \ll \Gamma_3$, then the transmission is limited
by the contact to the Fermi liquid.
The resonances are then insensitive to the Luttinger
liquid correlations, and have the Fermi liquid form (11).

(2) {\it $\nu=1/3$ dominated sequential tunneling} 
Since $\Gamma_3$ decreases as the temperature
is lowered, while $\Gamma_1$ remains temperature independent,
it is possible that the two could cross at a temperature
above $\Gamma_1$.  In this case the system would remain in the
sequential tunneling regime, but the tunneling would be limited
by $\Gamma_3$.  Provideded the coupling to $\nu = 1/3$ is 
sufficiently weak that $\Delta_3$ has not renormalized
significantly then the resonances peaks will vary as $T$
and the lineshape will be given by (12).  This will be the
case provided $t_3 \ll 1$, or equivalently
$\Gamma_3(T) \ll \tau_c T^2$, where $\tau_c^{-1}$ is the
high energy cutoff.   

(3) {\it Renormalized sequential tunneling}
For $\tau_c T^2 \ll \Gamma_3 \ll T$, $\Gamma_1 \ll T$,
one is no longer in the weak coupling limit.  Nonetheless,
as shown in Fig. 1, for $t_3 < t_3^*$ and $t_1=0$
the system flows to a fixed point with $t_3 = 0$. 
In this limit, $\Gamma_3 \ll T$,
so tunneling is sequential, however, the exponents
$\Delta_1$ and $\Delta_3$ are renormalized:
$1<\Delta_3<3/2$, and $1/2<\Delta_1<2 - \Delta_3$.
In this regime, the resonances can be described by
combining (5), (8) and (9).
The peak heights will vary as $T^\alpha$.  If  $\Gamma_1 < \Gamma_3$
then $-1 < \alpha < 0$.  If $\Gamma_1 > \Gamma_3$ then
$0 < \alpha < 1$.

(4) {\it Strong Coupling Regime}
When $T < \Gamma_1$ or $\Gamma_3$, the system is in a regime
where the impurity is strongly coupled to the leads.  This
regime will be characterize by the zero temperature phase
diagram shown in Fig. 2.  Generically, the impurity will
be strongly coupled to either the Fermi liquid or the
$\nu = 1/3$ lead.  In that case, the conductance will
have the ``off resonance" $T^2$ temperature dependence.
However, if the coupling $t_3$ is sufficiently strong, 
and the couplings are such that the system is
precisely on the phase boundary, then the system will
be at a perfect resonance.   Achieving the perfect resonance
requires two parameters to be tuned: the resonance energy
$\varepsilon_0$ and the coupling to $\nu=1/3$, $t_3$.
At the resonance the peak conductance should increase as the
temperature is lowered, approaching a peak value of 
$e^2/2h$, while the width becomes narrower, ultimately
having the universal lineshape given in (17) and (18).  
  
Recently, in their cleaved edge overgrowth experiments,
Grayson et al. have observed resonances 
in tunneling between a metal and the edge of a 
$\nu = 1/3$ quantum Hall state as a function of magnetic field\cite{grayson2}.  
These resonances have a peak conductance which increases
as approximately $1/T$ as the temperature is lowered and a line shape that is
well fit by the derivative of the Fermi function.  Since
the peak height of the resonances is quite small
(of order $.005 e^2/h$ at 25 mK), it is most likely that 
in this temperature range the system is in Fermi liquid dominated 
sequential tunneling regime.
From the peak height at 25 mK, one can estimate
$\Gamma_1 \approx 0.1 {\rm mK}$, which also gives a lower bound
on $\Gamma_3$, so that $0.1 {\rm mK} < \Gamma_3 < 25 {\rm mK}$.  

In these samples signatures of Luttinger liquid behavior in the
resonances may not occur until much lower temperatures,  
and reaching the strong coupling regime may prove difficult.
Nonetheless it would be interesting to probe
such resonances either by going to lower temperature or by
somehow increasing the coupling to the impurity state.
A second tuning parameter, such as a gate
voltage would allow additional flexibility in controlling 
$\Gamma_1$ and $\Gamma_3$, and would allow for the possibility
of tuning to the perfect resonance.  
The first signature of strong coupling behavior
would be a deviation from the $1/T$ dependence of the 
peak heights of the resonances.

\acknowledgements

It is a pleasure to thank C. de C. Chamon, E. Fradkin and
S. Sondhi for helpful discussions and A. Chang, M.
Grayson and D. Tsui for sharing their experimental
results prior to publication.  This work has been supported
by the National Science Foundation under grant
DMR 95-05425.

\end{document}